# Nearby Seyfert galaxies are possible sources of cosmic rays above $4\times10^{19}$ eV: UPDATED RESULTS

A. V. Uryson

Lebedev Physics Institute of Russian Academy of Sciences, Moscow 117924, Russia

**Abstract**—Arrival directions of 63 extensive air showers with energies $4\cdot10^{19}<E\leq3\cdot10^{20}$ eV detected by the AGASA, Yakutsk, Haverah Park arrays are analyzed in order to identify possible sources of cosmic rays with these energies. We searched for active galactic nuclei within error boxes around the shower-arrival directions and calculated the probabilities of objects being in the error boxes by chance. Our previous result obtained in 1996-2001 (see references in the text) is confirmed: the probabilities are small, $P>3\sigma$ ($\sigma$ is the parameter of Gaussian distribution) for Seyfert galaxies with redshifts $z<0.01$ (along with BL Lacertae objects). The Seyfert galaxies are characterized by moderate luminosities ($L<10^{46}$ erg/s) and weak radio and X-ray emission.

**1 Introduction**

Particles initiating extensive air showers with energies $E>4\cdot10^{19}$ eV are likely to have an extragalactic origin (see e.g. [1, 2]). In this case, the spectrum of extragalactic cosmic rays may abruptly steepen near $10^{20}$ eV due to their interaction with the microwave background radiation [3, 4]. The UHECR data do not show Greisen-Zatsepin-Kuzmin (GZK) cutoff [5]. GZK cutoff should not be observed if sources of ultra high energy cosmic rays (UHECR) are relatively nearby objects: the mean free path of particles with energies $E<10^{20}$ eV in the background radiation field is about ~40-50 Mpc, and particles with energies up to $E\approx10^{21}$ eV should traverse distances of about 10-15 Mpc essentially unattenuated [6].

UHECR sources considered in the literature can be divided into three categories. The first includes astrophysical objects, such as pulsars, active galactic nuclei, the hot spots and cocoons of powerful radio galaxies and quasars, gamma-ray bursts, and interacting galaxies ( [7] and references therein). The second category of proposed UHECR sources is cosmic topological defects [8], and the third is the decay of supermassive metastable particles of cold dark matter that have accumulated in galactic halos [9]. Direct identification of astrophysical objects is possible only in the first case. In other cases, any objects falling within error boxes centered on particle arrival directions should be chance coincidences. In our previous papers [10-12] we searched for possible sources within error boxes centered on the arrival directions of showers, and calculated the probabilities of objects being in the error boxes by chance. We found this probability to be rather small, $P>3\sigma$ for Seyfert galaxies with redshifts $z<0.01$, i.e. located within 40 Mpc ($H = 75$ km s$^{-1}$ Mpc$^{-1}$). In [12] we found that $P>3\sigma$ also for Blue Lacertae objects (BL Lac). For pulsars and radio galaxies the probability $P$ is about 0.1. The result that Bl Lac's are probable sources of UHECRs are also got in [13]. Other results on UHECR sources identification are the following. Quasars are found to be UHECR sources in [14]. However this result was discussed in [15], and it was shown that neither compact radio sources nor gamma-ray emitting blazars seemed to be sources of UHECRs.

In this report we performed identification of UHECR sources with larger statistics of showers and with the catalogue [16] to search for the sources.

**2 The identification procedure**

We used showers whose arrival directions were published along with their errors. For identification procedure, we selected showers with errors in arrival directions ($\Delta\alpha$, $\Delta\delta$)$\leq 3^0$ in equatorial coordinates: 58 events with $E>4 \cdot 10^{19}$ eV [8], 4 Yakutsk showers with $E>4 \cdot 10^{19}$ eV [17], their errors were computed in our previous paper [11], and 1 Haverah Park showers with $E \geq 10^{20}$ eV [18], its error was computed in [14].

Different objects occur around the patricle arrivals. We use the following procedure to obtain the probability of a chance occurence of objects near shower arrivals. The showers were subdivided into several groups depending on Galactic latitude $b$ of arrival directions, and in each error box we looked for objects of the given type. We counted the number of showers $K$ in each group and the number of showers $N$ which have at least one object of the given type within the error box. The showers were subdivided in Galactic latitude $b$ in order to exclude events clearly lying in the galaxy 'avoidance zone'. We calculated the probabilities that objects of the given type would fall in the fields of search of $N$ of the total of $K$ showers by chance as follows. Showers with randomly distributed arrival coordinates were simulated. The coordinates of the simulated showers were determined by a random-number generator [19] within a survey band $\alpha = 0-24$ hr and $\delta = -10-90°$. We subdivided simulated showers into groups in the same way as real showers. Each simulated group contains the number $K$ of showers equal to those observed. We then counted in each simulated group the number of showers $N_{sim}$ having at least one object of the given type located in the error box ($N_{sim}$ can take values in the interval $N_{sim} \leq K \leq N$). In a group of $K$ showers, the probability $P$ of a chance occurrence of galaxies in the field of search of a given number of showers $N_{sim}$ was determined as $P = \Sigma(N_{sim})_i/M$ where $i$ is ranging from 1 to $M$, $M=10^5$ is the number of trials performed for each group. By the law of probability the coincidence is by chance if the probability $P$ is $P>3\sigma$, where $\sigma$ is the parameter of Gaussian distribution.

What is the size of the region of search? Statistics and the law of probability give the following data [20]: the probability that the particle coordinates are within the one-mean-square error box is 68 per cent, the probability for the two-mean-square error box is 95 per cent, and for the three-mean-square error box it is 99.8 per cent. It means that more than 30 per cent of the objects are excluded from the analysis a priori using 1-error box, only 5 per cent of objects are lost with 2-error box, and essentially all objects are considered with 3-error box. Using 2-error box is less strict than using 3-error box, and it is more accurate as compared with using 1-error box. One believes that using 1-error box region would reduce the chance coincidence against 2- and 3-error boxes.

In the papers [13-15] 1-error box was used. Here we use both 1-, 2-, and 3-error boxes in the identification procedure.

The optical coordinates of the galaxies and pulsars are accurate to several arcseconds, and fields of search were determined solely by the errors in the shower coordinates.

## 3 Results

The catalogue [16] contain both Seyferts with detailed classification and objects which are probably or possibly Seyferts, because of a lack of avaliable data. We found the probabilities of chance coincidence in two cases: for all Seyferts and for Seyferts with detailed classification.

For all nearby Seyferts the probabilities of chance coincidence using 1-, 2-, and 3-error boxes $P_1(N)$, $P_2(N)$, and $P_3(N)$ are the following ($N$ is the number of showers having at least one nearby galaxy within the error box; if some showers in a group had galaxies in regions smaller than 3-error box but larger than 2-error box, we determined the weighted mean error box, so 2-error box means also 2.1- or 2.2-error box, and similarly 1-error box means also 1.2- or 1.3-error box):

63 showers with no selection in Galactic latitude, $P_1(16)=1.1 \cdot 10^{-3}$, $P_2(27)=3.6 \cdot 10^{-4}$, $P_3(29)=2.4 \cdot 10^{-2}$;

54 showers with $|b|>11.2^0$ $P_1(16)=1.2 \cdot 10^{-3}$, $P_2(26)=6.5 \cdot 10^{-4}$, $P_3(29)=1.8 \cdot 10^{-2}$;

37 showers with $|b|>21.9^0$ $P_1(13)=3.2 \cdot 10^{-3}$, $P_2(23)=1.8 \cdot 10^{-4}$, $P_3(23)=2.5 \cdot 10^{-2}$;

27 showers with $|b|>31.7^0$ $P_1(14)=5.1 \cdot 10^{-4}$, $P_2(23)=2.0 \cdot 10^{-5}$, $P_3(23)=9.5 \cdot 10^{-3}$.

Here probabilities are $P>3\sigma$ for 1- and 2- error boxes for showers at any latitudes, except $|b|>21.9^0$, where $P_1 \approx 2.95\sigma$.

For nearby Seyferts having detailed classification, the probabilities are:

63 showers with no selection in Galactic latitude, $P_1(12)=1.1 \cdot 10^{-2}$, $P_2(23)=3.2 \cdot 10^{-3}$, $P_3(27)=3.2 \cdot 10^{-2}$;

54 showers with $|b|>11.2^0$ $P_1(12)=1.5 \cdot 10^{-2}$, $P_2(22)=5.2 \cdot 10^{-3}$, $P_3(27)=2.3 \cdot 10^{-2}$;

37 showers with $|b|>21.9^0$ $P_1(9)=3.0 \cdot 10^{-2}$, $P_2(19)=3.0 \cdot 10^{-3}$, $P_3(21)=3.7 \cdot 10^{-2}$;

27 showers with $|b|>31.7^0$ $P_1(10)=1.0 \cdot 10^{-2}$, $P_2(19)=1.1 \cdot 10^{-3}$, $P_3(21)=2.2 \cdot 10^{-2}$.

Here probabilities are $P \geq 3\sigma$ for 2-error boxes at any latitudes, except showers at $|b|>11.2^0$ where $P_2 \approx 2.80\sigma$.

Because of low values of probabilities it is difficult to ignore nearby active galaxies as possible UHECR sources. (Probabilities increase with increasing $z$, see [10] for $P(z)$ relations for showers arriving from sky areas located at arbitrary $b$.)

For comparison, we get large probabilities, $P \sim 0.01\text{-}0.1$ using the same procedure for identifying radio galaxies.

Acceleration of particles up to $10^{21}$ eV in moderate Seyfert nuclei is described in [21].

## Acknowledgments

I am indebted to Profs. V.L. Ginzburg, V.S. Berezinsky, N.S. Kardashev, B.V. Komberg, O.K. Sil'chenko, and A.V. Zasov for discussions of this work at its different stages.